\documentstyle[11pt,epsfig]{article}
\begin{document}
\topmargin -1.4cm
\oddsidemargin -0.8cm
\evensidemargin -0.8cm
%%-------------------------------------------------------
%%%%%%%%%%%%%%%%%%%%%%%%%%%%%%%%%%%%%%%%%%%%%%%%%%%%%%%%%
%%-------------------------------------------------------
\def\g#1{{\scriptscriptstyle (\! #1 \! )}}
\def\DELTA#1{{
     \!\begin{array}{c}\setlength{\unitlength}{.5 pt}
     \begin{picture}(35,25)
        \put(15, 0){\line(0,1){10}} \put(20, 0){\line(0,1){10}}
        \put(15, 0){\line(1,0){ 5}} \put(15,10){\line(1,0){ 5}}
        \put(15,25){\line(1,0){5}}  \put(15,15){$\scriptstyle {#1}$}
        \put(15,15){\oval(20,20)[l]}\put(20,15){\oval(30,20)[r]}
     \end{picture}\end{array} \!
}}
\def\THETA#1#2#3{{
     \begin{array}{c}\setlength{\unitlength}{.5 pt}
     \begin{picture}(40,40)
        \put(18,32){$\scriptstyle {#1}$}
        \put( 0,15){\line(1,0){40}} \put(18,17){$\scriptstyle {#2}$}
        \put(20,15){\oval(40,30)}   \put(18, 2){$\scriptstyle {#3}$}
        \put( 0,15){\circle*{3}}    \put(40,15){\circle*{3}}
     \end{picture}\end{array}
}}
\def\TET#1#2#3#4#5#6{{
     \begin{array}{c}\setlength{\unitlength}{.8 pt}
        \begin{picture}(50,30)
        \put( 0,15){\line(1,-1){15}} \put(0,22){${\scriptstyle {#2}}$}
        \put( 0,15){\line(1, 1){15}} \put(0, 0){${\scriptstyle {#1}}$}
        \put( 0,15){\circle*{3}}
        \put(30,15){\line(-1, 1){15}} \put(28,20){${\scriptstyle {#4}}$}
        \put(30,15){\line(-1,-1){15}} \put(28, 2){${\scriptstyle {#5}}$}
        \put(30,15){\circle*{3}}
        \put( 0,15){\line(1,0){30}} \put(12,16){${\scriptstyle {#6}}$}
        \put(15,30){\line(1,0){25}} \put(15,30){\circle*{3}}
        \put(15, 0){\line(1,0){25}} \put(15, 0){\circle*{3}}
        \put(40, 0){\line(0,1){30}} \put(42,12){${\scriptstyle {#3}}$}
    \end{picture}\end{array}
}}

%%%%%%%%%%%%%%%%%%%%%%%%%%%%%%%%%%%%%%%%%%%%%%%%%%%%%%%%%%%%%%%%
%%%%%%%%%%%%%%%%%%%%%%%%%%%%%%%%%%%%%%%%%%%%%%%%%%%%%%%%%%%%%%%%
%%%%%%%%%%%%%%%%%%%%%%%%%%%%%%%%%%%%%%%%%%%%%%%%%%%%%%%%%%%%%%%%
%%%%%%%%%%%%%%%%%%%%%%%%%%%%%%%%%%%%%%%%%%%%%%%%%%%%%%%%%%%%%%%%

\title{\Large{{\bf Canonical ``Loop'' Quantum Gravity and Spin
Foam Models}}}

\vspace{2.5cm}

\author{~\\{\sc R. De Pietri$^{(1)(2)}$}\\~\\
(1) Dipartimento di Fisica\\Viale delle Scienze, I-43100 Parma, Italy\\
and INFN Gruppo collegato di Parma\\~\\
(2) Centre de Physique Th\'eorique CNRS\\
Case 907 Campus de Luminy\\F-13288 Marseille Cedex 9, France}
\date{March 19, 1999} 
\maketitle
%%%\vfill
\begin{abstract}
The canonical ``loop'' formulation of quantum gravity is a mathematically 
well defined, background independent, non perturbative standard 
quantization of  Einstein's theory of General Relativity.  
Some among the most meaningful results of the theory are: 
1) the complete calculation of the spectrum of geometric quantities
like the area and the volume and the consequent physical predictions
about the structure of the space-time at  the Planck scale; 
2) a microscopical derivation of the Bekenstein-Hawking black-hole
entropy formula. Unfortunately, despite recent results,
the dynamical aspect of the  theory (imposition of the Wheller-De Witt 
constraint) remains elusive.

After a short description of the basic ideas and the main results 
of {\it loop} quantum gravity we show in which sence the 
exponential of the super Hamiltonian constraint leads to 
the concept of {\it spin foam}
and to a four dimensional formulation of the theory. Moreover, we show
that some topological field theories as the BF theory in 3 and 4
dimension admits a spin foam formulation. We argue that 
the spin-foams/spin-networks formalism it is the natural framework
to discuss loop quantum gravity and topological field theory.  

\end{abstract}

\vfill
\begin{flushleft}
CPT-99/P.3796 \\
University of Parma Preprint UPRF-99-01 \\
{\it xxx-archive:} gr-qc/9903076.\\[1cm]
To appear in the proceeding of the XXIII Congress
of the Italian Society for General Relativity and
Gravitational Physics (SIGRAV), Monopoli (ITALY), 
September $21^{st}$-$25^{th}$, 1998.
\end{flushleft}

\thispagestyle{empty}
\newpage

\setcounter{page}{2}
\setcounter{equation}{0}

%%%%%%%%%%%%%%%%%%%%%%%%%%%%%%%%%%%%%%%%%%%%%%%%%%%%%%%%%%%%%%%%%%
%%
%%           S E C T I O N
%%
%%%%%%%%%%%%%%%%%%%%%%%%%%%%%%%%%%%%%%%%%%%%%%%%%%%%%%%%%%%%%%%%%%

\section{Introduction}

The {\it loop approach} to quantum gravity
\cite{Rovelli:1988,Rovelli:1990} has reached a mature status as a
physical theory and it is one of the most likely candidate to be 
a complete consistent quantum theory of gravitational phenomena (for a
detailed review and update bibliography we refer to
\cite{Rovelli:1998a} for quantum gravity in general and to
\cite{Rovelli:1998b} for the loop approach in particular).  
In this approach it was possible:
%%%%
{\it i)} to construct the auxiliary Hilbert space
\cite{Ashtekar:1992,Ashtekar:1994} appropriate to the discussion of
the kinematics
%%%%
{\it ii)} to find an explicit basis (the spin-networks basis) on this
Hilbert space \cite{Baez:1994,Rovelli:1995};
%%%% 
{\it iii}) to rigorously solve the 3-diffeomorphism constraints
\cite{Ashtekar:1995};
%%%%
{\it iv)} to construct operators corresponding to volume and area and
determine their spectrum
\cite{Rovelli:1995a,DePietri:1996,Ashtekar:1997,Ashtekar:1998a}.
%%%%
{\it v}) to derive the Beckenstein-Hawking formula for the entropy of
Black Holes \cite{Rovelli:1996a,Barreira:1996,Krasnov:1997,Ashtekar:1998}

The elusive feature of the loop approach to quantum gravity is 
that in the {\it frozen-time} canonical picture the relation 
between operator and physical observable is cumbersome. A space 
time picture  was lacking until M. Reisenberger a 
C. Rovelli \cite{Reisenberger:1997} derived a spin foam model
(using the terminology introduced by J. Baez \cite{Baez:1998}), 
to which they refer as {\it world sheet model}, from the Exponential 
of the super-Hamiltonian constraint. They also show that this
is formally related to the construction of the 4 dimensional 
path integral for gravity.  This work was a first step in 
the direction of developing a sum over surfaces framework 
in ``loop quantum gravity'' whoose importance was first claimed  
by Reisenberger in \cite{Reisenberger:1994}.
This recent developments showed surprising formal analogy between loop
quantum gravity and other approaches to the quantization of gravity. 
The model of dynamical triangulation \cite{Ambjorn:1997} and the
Ponzano-Regge model\footnote{This model correspond to three dimensional 
BF theory \cite{Ooguri:1991,Ooguri:1992a,Archer:1991}
and it is closely related  to Witten's Chern-Simons approach 
to the quantization of 2+1 dimensional gravity 
\cite{Witten:1988,Witten:1988a,Witten:1988c,Witten:1989,Witten:1989a}} 
\cite{Ponzano:1968} are exactly models of this
kind. The analogy is even deeper if one considers the
Ponzano-Regge model  [the physicist version of the celebrated 
Turaev-Viro state sum  model for 3-manifold invariant
\cite{Turaev:1992})] that can be formulated exactly as a sum 
over  branched surfaces with colored graph observables (spin networks)
on the boundary.

The previous consideration, together with the fact that the topological 
BF  field theory in 4 dim \cite{Ooguri:1992b}, 
\cite{Crane:1997} (see also \cite{Carter:1998}) admits an analogous 
formulation as a spin foam model [in this case,the spin foam 
dual to a simplicial decomposition (triangulation)], strongly suggest that 
quantum gravity should be naturally formulated
using the language of spin foam models. Now,
two explicit proposals for the realization of this idea have been 
put forward. The first one (by Barret and Crane
\cite{Barrett:1998}) is related to the discretization of the path integral
of the Euclidean Plebanski action
\cite{Plebanski:1977} (see \cite{DePietri:1998,Reisenberger:1998a})
and the second (by M. Reisenberger \cite{Reisenberger:1997b}) of 
self dual Euclidean gravity \cite{Reisenberger:1995}. The last
model is the explicit  quantization of the 
lattice model \cite{Reisenberger:1997a} that 
has self dual Euclidean gravity as its continuum limit.
This research line is deeply connected to the idea of express 
classical gravity as a deformation (by the imposition of
additional Lagrange constraints) of a topological BF field theory
4 dimension. The idea is that  \cite{Horowitz:1989} topological BF 
theory, in the context of diff-invariant theory, can play a role 
analogue to the one of free theory in standard Poincar\'e invariant  %%
quantum fields theories. This idea is at the origin of the
systematic proposal for deriving {\it Spin Foam Weight Factors}
from a Classical Action Principle that has been recently 
proposed  by Freidel and Krasnov \cite{Freidel:1998}. 

The discussion here it is in framework of piecewise linear topology and we
refer the reader to \cite{Rourke:1972} for details. Intuitively, this
setting is equivalent to assume that  cells, surfaces and edges
that we use to decompose the manifold are finite unions of
piecewise analytic sub-manifolds. In particular, a graph is
a finite collection of analytic path.  This framework as certain
advantages and the resulting theory is simpler than the ones in the
analytic or in the smooth setting.  The utility 
of the PL setting in loop quantum gravity was first pointed out by
Zapata \cite{Zapata:1997a,Zapata:1997b}. All the topological
objects (graph, manifold, cellular decomposition, ....) are assumed
orientable.  For the Wigner nJ-symbols we
adopt the non normalized graphical convention of
\cite{Kauffman:1994,DePietri:1997} and we restrict to the case of the
group $SU(2)$ and use Penrose's binor formalism
\cite{Penrose:1971a}. Most of the formulas remain valid in the quantum
group case, but, in this case, factors related to the braiding can 
appear.

%%%%%%%%%%%%%%%%%%%%%%%%%%%%%%%%%%%%%%%%%%%%%%%%%%%%%%%%%%%%%%%%%%
%%
%%           S E C T I O N
%%
%%%%%%%%%%%%%%%%%%%%%%%%%%%%%%%%%%%%%%%%%%%%%%%%%%%%%%%%%%%%%%%%%%

\section{Loop Quantum Gravity}

Loop quantum gravity is based on the Hamiltonian formulation of
classical general relativity in terms of the canonically conjugated
variables $\tilde{E}^a_i(x)$ (densitized triad fields) and $A_a^i(x)$
(Ashtekar's connection) \cite{Ashtekar:1986}.  The dynamics is given,
as it happens for any diff-invariant theory, only in terms of constraints. 
In our case:
\[
\begin{array}{lr}
C_i(x)= {\cal D}_a\tilde{E}^a_i(x)\simeq 0  
&\mbox{(Gauge constraints)}\\
{\cal H}_a(x)= \tilde{E}^b_i(x) F^i_{ab}(x)\simeq 0  
&\mbox{(Diffeomorphism constraints)}\\
{\cal H}(x)= \frac{\epsilon^{ijk}}{\sqrt{g}}
             \tilde{E}^a_i(x)\tilde{E}^b_j(x) F^k_{ab}(x)
          \simeq 0  
&\mbox{(Hamiltonian constraints).} 
\end{array}
\]

The loop program does the quantization of gravity in a
straightforward way. It is based on the assumption that a particular
class of classical functions (the ${\cal T}$ observables, i.e., the
{\it traces} of the {\it holonomy} of Ashtekar's connection) become
well defined quantum operators. I.e., it is 
assumed that, for any choice of the close loop $\gamma$, to the  
classical  functional of the Ashtekar connection:
\begin{equation}
{\cal T}_\gamma[A_a^i(x)] = {\rm Tr}\left[ \exp\left(
 \int_\gamma A_a(x) dx^a\right) \right]
\end{equation}
corresponds a well defined quantum operator on a Hilbert
space ${\cal H}$.  All the other properties should be
essentialy derived from this assumption. 
A consequence of this choice is that the canonical variables 
$\tilde{E}^a_i(x)$,$A_a^i(x)$ do not become well defined 
operator on the Hilbert space ${\cal H}$ and the previous definition
of the constraint (unless properly regularized) are meaningless on 
${\cal H}$. This is one of the main technical reasons behind the 
difficulty of dealing with the super-Hamiltonian constraint.

Given this assumption, there are two possible ways to define the
theory. Explicitly construct the vector space on which
these operators are well defined (loop representations, see
\cite{DePietri:1996}), in a algebraic way, and then determine the 
correct scalar product. Alternatively, construct a
Hilbert space ${\cal H}$ (the connection representation) 
and show that, on this Hilbert space, these operators
are well defined \cite{Ashtekar:1995}.  These two approaches are
equivalent \cite{DePietri:1997}.  However, two technical but important
decision must be taken.  The first one is connected to the properties
of the loops $\gamma$. The request of piecewise analyticity
brings to a different setting with respect to the smooth case since two
piecewise analytic loop can have only a finite number of isolated
intersection points. The second is that these constraction 
are well defined only in the case of a real connection.
Now, the standard Ashtekar connection is naturaly real only in 
the case of Euclidean gravity while in the case of Lorentzian 
gravity is naturally complex and indeed what follows 
naturally applies only to Euclidean gravity.
However, it is important to note that, it is possible to use a 
real connection in the Lorentzian case too (the Barbero connection
\cite{Barbero:1995,Barbero:1996}) at the prize of a more complicated
form for the super-Hamiltonian constraint.

\subsection{The auxiliary Hilbert space}

A full discussion  of the kinematical setting is not important here.
We refer to \cite{Ashtekar:1995,DePietri:1997} 
for the technical details and for complete references.
For the following discusson it is sufficient to recall the
basic structure of  the Hibert space ${\cal H}$.
Since a connection associates a group element 
$g_{e_i} ={\cal P}\exp(-\int_{e_i} A)$ to
each edge (segment embedded in $M^3$) of the graph $\gamma$.
It is natural to consider for each graph $\gamma$ the Hilbert space 
(${\cal H}_\gamma=L^2[G^n,d\mu(g)]$) of square integrable function, 
with respect to the unique normalized Haar measure $d\mu(g)$, 
on $n$ (the number of edges ${e_i}$ in $\gamma$) copies of the 
group [$G=SU(2)$]  $G^n$.  ${\cal H}$ can be characterized 
as the projective limit of the previous family of  Hilbert 
spaces ${\cal H}_\gamma$ 
associated to all possible graphs $\gamma$ embedded into $M^3$.  
%%%%%%%%%%FIG 1%%%%%%%%%%%%%%%%%%%%%%%%%%%%%%%%%%%%%%%%%%%%%%%%%%%%%
\begin{figure}[t]
\begin{center}
\mbox{\epsfig{file=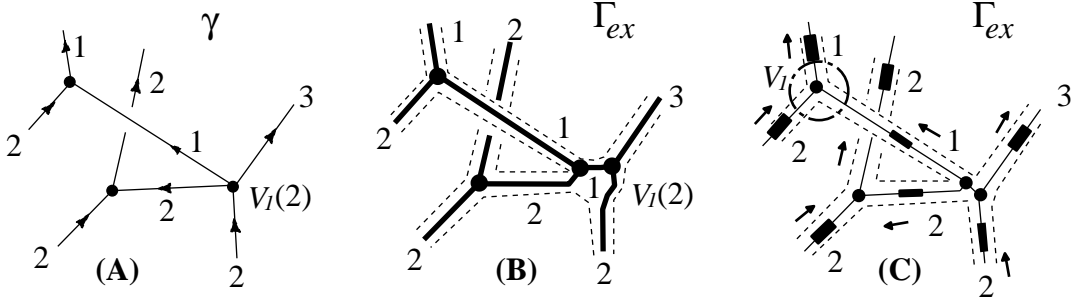}} 
\end{center}
\caption{The graph $\gamma$, a possible extended-planar 
projection $\Gamma_{ex}$ and the graphical representation
of a spin-network cylindrical function}
\label{fig:graph}
\end{figure}
%%%%%%%%%%%%%%%%%%%%%%%%%%%%%%%%%%%%%%%%%%%%%%%%%%%%%%%%%%%%%%%%%%
The gauge invariant Hilbert space ${\cal H}_{\rm
aux}$ is the projective limit of the gauge invariant sub-sector
of the Hilbert spaces ${\cal H}_\gamma$. Moreover, the Peter-Weyl
theorem gives us a natural basis in this space of gauge invariant
functions $f_\gamma(g_{e_1},\ldots,g_{e_n})$, the spin networks basis:
\begin{equation}
f_\gamma(g_{e_1},\ldots,g_{e_n}) = \!\sum_{\vec{\pi},\vec{\iota}} \!
   f(\gamma,\vec{\pi},\vec{\iota}) 
  ~~{\cal T}_{\gamma,\vec{\pi},\vec{\iota}~}[A]
,~~~~
{\cal T}_{\gamma,\vec{\pi},\vec{\iota}}[A] \stackrel{def}{=}
  \left( \otimes_{i=1}^{\#^{\rm edge}} \pi_i(g_{e_i}) \right) \cdot
  \left( \otimes_{j=1}^{\#^{\rm vertex}} \iota_j \right) 
\label{eq:defSPINnet} \end{equation}
where: 
%%%%%
{\bf (i)} $\vec{\pi}=(\pi_1,\ldots,\pi_N)$ denotes the labeling of the
edges with irreducible representation $\pi_i$ of $G$;
%%%%%
{\bf (ii)} $\vec{\iota}=(\iota_1,\ldots,\iota_M)$ a labeling of the
vertices with invariant contractors $\iota_j$ (the intertwining
matrices $\iota_j$, in each of the vertices $v_j$, represents the
invariant coupling of the $n_j$ representations associated to the $n_j$
edges that start or end in $v_j$). In the case of loop 
quantum gravity (where $A_a^i$ is a $SU(2)$ connection) 
the vector of the basis (a spin network state) are  
characterized (see figure \ref{fig:graph}) by a 
colored (the irreducible representation of $SU(2)$ 
are label by the associated spin or by an integer,
that we call the color of the representations, that 
its twice the spin) graph. 

\subsection{The diffeomorphism invariant state}

A rigorous definition of the diffeomorphism invariant Hilbert
space can be obtained \cite{Ashtekar:1995} using the gauge averaging 
procedure. The essential idea is that on ${\cal H}_{\rm aux}$
there is a natural definition of the action of a 
diffeomorphism $\phi\in {\rm Diff}$ 
($\phi:M\rightarrow M$). In fact, in the spin-networks basis
we can associated to each diffeomorphism $\phi$ the operator  
\begin{equation}
 \hat{U}_\phi |S\rangle=\hat{U}_\phi |\gamma,\vec{\pi},\vec{\iota}~\rangle 
 = |\phi(S)\rangle = |S'\rangle =|\phi(\gamma),\vec{\pi},\vec{\iota}~\rangle ,
\end{equation}
where $S'=\phi(S)$ is the imagine under $\phi$ of the spin network $S$. 
At this point one can define the class of spin networks knots $s$ as the
equivalence class of spin networks under diffeomorphism, i.e.
$S,S'\in s$ it it exists a $\phi\in {\rm Diff}$ such that $S'=\phi(S)$.
In this way, we can define 
${\cal H}_{\rm Diff}= {\cal H}_{\rm aux}/{\rm Diff}$ with the 
scalar product:
\begin{equation}
\langle s|s' \rangle = N \int [D\phi] \langle S|\hat{U}_\phi S'\rangle
          = N \int [D\phi] \langle \phi(S)| S'\rangle
\label{sp:DIFF}
\end{equation}
where $N$ is an arbitrary, not yet fixed, normalization factor
and $S,S'\in {\cal H}_{\rm aux}$ are two arbitrary spin networks
in the $s$-knot class of $s$ and $s'$, respectively. The integration
in eq. (\ref{sp:DIFF}) is meaningful because the scalar product
of two  spin networks of ${\cal H}_{\rm aux}$ is different from zero 
only if they have support on the same graph. 

In the previous derivation of the Hilbert space ${\cal H}_{\rm Diff}$
we made an implicit assumption on the kind of diffeomorphism that
are allowed. Having previously assumed that the loops
are in the class of the piecewise analytics ones, it is natural 
to assume piecewise analytics diffeomorphism. In this case,
the quotient space ${\cal H}_{\rm Diff}$ is a separable Hilbert
space. This leads directly to the PL framework. We refer to 
the work of Zapata for the discussion of the subtleties involved
\cite{Zapata:1997b,Zapata:1997a}.

\subsection{The super-Hamiltonian constraint}

In the first presentation of {\it loop quantum gravity} 
\cite{Rovelli:1988,Rovelli:1990} was emphasized that loops without 
intersection, (i.e., spin network states whose graph $\gamma$ with
only bivalent vertices) it is automatically in the kernel
of the super-Hamiltonian 
constraint. This gives the knowledge of an infinite dimensional subspace 
of the {\bf physical} Hilbert space ${\cal H}_{{\rm Phys}}$.
Unfortunately, with the introduction of the volume operator 
\cite{Rovelli:1995a}, it was realized that all the spin network states based 
on graph $\gamma$ with all vertices of valence less that $3$
are in the kernel of the volume operator.  Consequently, all 
the previously mentioned states, belonging to 
the kernel of the super-Hamiltonian constraint, correspond to 
degenerate metrics and, therefore, can not correspond 
to genuine quantum 3 geometries.
This urged the definition of the super-Hamiltonian 
constraint ${\cal H}(x)$ on the whole diff-invariant 
Hilbert space ${\cal H}_{\rm Diff}$.  

The situation improved in 1996 
when Thiemann \cite{Thiemann:1996,Thiemann:1998a,Thiemann:1998b}
was able to construct an operator with the {\it naiv\"e} 
classical limit. Using this definition, it is possibile 
investigate the true {\bf physical} Hilbert space ${\cal H}_{{\rm Phys}}$.
For the purpose of the present discussion we do 
not need a complete account of its definition and its properties.
What it is interesting for us it is that: 1) it is possible to 
define such operator; 2) a rough idea of 
its action ${\cal H}_{\rm Diff}$. Its main property is that 
it acts locally around each vertex of the spin network. 
It adds or deletes an edge of color 1 
\begin{equation}
  \hat{A}_{vIJK\bar{\epsilon}\tilde{\epsilon}} = \left\langle 
  \begin{array}{c}\setlength{\unitlength}{.8 pt}
  \begin{picture}(65,50)
     \put( 0, 0){\line( 1,  1){20}} 
     \put( 0,20){\line( 1,  0){20}} 
     \put( 0,40){\line( 1, -1){20}}
     \put(20,20){\circle*{3}}
     \put(20,20){\line( 3,  2){30}} 
     \put(52,38){$\scriptstyle K$} 
     \put(20,20){\line( 3, -2){30}} 
     \put(52, 2){$\scriptstyle J$} 
     \put(45, 8){$\scriptstyle p$}  
     \put(45,30){$\scriptstyle q$} 
     \put(15, 2){$\scriptstyle p\!+\!\bar{\epsilon}$}  
     \put(15,32){$\scriptstyle q\!+\!\tilde{\epsilon}$}
     \put(35,10){\line(0,1){20}}
     \put(35,10){\circle*{3}} \put(35,30){\circle*{3}}
     \put(37,20){$\scriptstyle 1$}
  \end{picture}\end{array}
         \right| \hat{\cal H}[N] \left| 
  \begin{array}{c}\setlength{\unitlength}{.8 pt}
  \begin{picture}(65,50)
     \put( 0, 0){\line( 1,  1){20}} 
     \put( 0,20){\line( 1,  0){20}} 
     \put( 0,40){\line( 1, -1){20}}
     \put(20,20){\circle*{3}}
     \put(20,20){\line( 3,  2){30}} \put(52,38){$\scriptstyle K$} 
     \put(20,20){\line( 3, -2){30}} \put(52, 2){$\scriptstyle J$} 
     \put(45, 8){$\scriptstyle p$}  \put(45,30){$\scriptstyle q$} 
  \end{picture}\end{array}
    \right\rangle
\end{equation}
with well definite weight factors 
[($\hat{A}_{vIJK\bar{\epsilon}\tilde{\epsilon}}$) matrix elements of
the operator] that can be explicitly computed \cite{Borissov:1997}.

\subsection{The hierarchy of Hilbert spaces of loop quantum gravity}

Summarizing, the mathematical definition of the loop quantization of 
gravity, and the determination of the {\bf physical} Hilbert 
space ${\cal H}_{{\rm Phys}}$, involves the analysis of the following 
hierarchy of Hilbert spaces: 
\begin{equation}
{\cal H}               ~\stackrel{C_i(x)}{\Longrightarrow}~
{\cal H}_{{\rm aux}}   ~\stackrel{{\cal H}_a(x)}{\Longrightarrow}~
{\cal H}_{{\rm Diff}}  ~(?)\!\stackrel{{\cal H}(x)}{\Longrightarrow}\!(?)~
{\cal H}_{{\rm Phys}}
\end{equation}
where an arrow means imposition of a constraint and the 
questions mark emphasis the fact that, the last step, it is still not
completely understood. In fact: 1) the physical correctness of the 
Thiemann's proposal for the regularization of the super-Hamiltonian
constraint has been questioned; 2) more of one variant of this
operator can be construct and indeed the problem of determine the
correct one; 3) despite \cite{Thiemann:1998c} there is not a clear 
understanding of the kernel of the super-Hamiltonian constraint nor
of the class of operator that are well defined on this Hilbert space.

%%%%%%%%%%%%%%%%%%%%%%%%%%%%%%%%%%%%%%%%%%%%%%%%%%%%%%%%%%%%%%%%%%
%%
%%           S E C T I O N
%%
%%%%%%%%%%%%%%%%%%%%%%%%%%%%%%%%%%%%%%%%%%%%%%%%%%%%%%%%%%%%%%%%%%

\section{The exponential of the Hamiltonian constraint}

Reisenberger and Rovelli in \cite{Reisenberger:1997,Rovelli:1998c}
suggested (taking advance of the existence of a well defined version
of the Hamiltonian constraint
\cite{Thiemann:1996,Thiemann:1998a,Thiemann:1998b} on ${\cal H}_{{\rm
Diff}}$) that the better way to address this last remaining problem is
to use a 4-dimensional picture.  In the case of classical
general relativity  the lapse ${N}(x,\tau)$ and shift function
$N^a(x,\tau)$ ($N^\mu(x,\tau)=d\Sigma^\mu(x,\tau)/d\tau$) generate the
evolution along a foliations ($\Sigma(\tau)$) between $\Sigma(0)$
and $\Sigma(1)$. In an analogous way, in the quantum case, 
we can say that the lapse and shift function 
generate the evolution from a spin network state 
$|s,\Sigma(0)\rangle \in {\cal H}_{\rm aux}$ associated to the
boundary $\Sigma(0)$ to the spin network state
$|s',\Sigma(1)\rangle\in {\cal H}_{\rm aux}$ associated to the
boundary $\Sigma(1)$.
Formally\footnote{One should keep in mind
that the following formula do not have a well definite meaning: there
are no rigorous or unambiguous definitions of the 3-diff constraints nor of the
Hamiltonian constraints in the Hilbert space ${\cal H}_{aux}$.} 
we have
\begin{eqnarray}
\frac{d}{dt} |S',\Sigma(1)\rangle
   &=&  ~\bigg( N(x,\tau)\hat{\cal H}(x)+N^a(x,\tau)\hat{\cal H}_a(x)
         \bigg)
  |S,\Sigma(t)\rangle
\\
|S',\Sigma(1)\rangle &=&    {\cal T}\!\!\exp\bigg[\!\!-\!{\rm i}
         \!\int_0^T \!\!\! d\tau\!\int\!\! d^3\!x
        ~\bigg( N(x,\tau)\hat{\cal H}(x)+N^a(x,\tau)\hat{\cal H}_a(x)
         \bigg)\bigg]
  |S,\Sigma(0)\rangle
~~.
\end{eqnarray}
The last equation suggests to define the transition amplitude,
averaged over all the possible embedding (bordism) interpolating from
$\Sigma(0)$ and $\Sigma(1)$:
\begin{eqnarray}
&&Z[S',S;1] = \langle S',\Sigma(1)|S,\Sigma(0)\rangle =
\label{def:PI}\\
&&~~~~~= \int [dN(x,\tau)][dN^i(x,\tau)] \langle s'| {\cal
   T}\!\!\exp\bigg[\!\!-\!{\rm i} \!\int_0^1 \!\!\! d\tau\!\int\!\!
   d^3\!x ~\bigg( N(x,\tau)\hat{\cal H}(x)+N^a(x,\tau)\hat{\cal
   H}_a(x) \bigg)\bigg] |S\rangle ~~.  \nonumber
~~,
\end{eqnarray}
where the integrations $[dN(x,\tau)][dN^i(x,\tau)]$ are over all the
lapse and shift functions that represent a bordism between $\Sigma(0)$
and $\Sigma(1)$. They note that it is possible to give a
well defined meaning to the integration over the lapse of the
exponential of infinitesimal diffeomorphism in term of the gauge
averaging procedure used in \cite{Ashtekar:1995} to solve the
diffeomorphism constraints:
\begin{eqnarray}
  \hat{U}_{\phi_{N^i}} &=:& \exp\bigg[\!\!-\!{\rm i} \int \!  d^3\!x ~
                     N^a(x)\hat{{\cal H}}_a(x) \bigg]
\label{def:expD} \\
  |s \rangle &:=& \int [D\phi] ~ \hat{U}_\phi |S\rangle =: \int
              [dN^a(x)] ~ \hat{U}_\phi |S\rangle
\label{def:DIs}
\end{eqnarray}
and then, using a series of formal argument, that the definition
(\ref{def:PI}) depends only on the s-knot classes $s$ and $s'$
of $S$ and $S'$, respectively. Morover, they show that it can be
expressed in term of a diff-invariant definition (as the one 
proposed by Thiemann) of the super-Hamiltonian constraint 
and of its associated proper time propagator:
\begin{equation}
Z_{RR}[s',s;1] = \int dT <s'|e^{{\textstyle -{\rm i}}~T~{\cal H}}|s> =
          \int dT \sum_\sigma \frac{({\rm
          i}~T)^{n(\sigma)}}{n(\sigma)} \prod_{({\rm
          coloring~of~}\sigma)} \prod_{v\in[\sigma]} A_v(\sigma)
\label{RR}
\end{equation}
%%%%%%%%%%%%%%%%%%%%%%%%%%%%%%%%%%%%%%%%%%%%%%%%%%%%%%%%%%%%%%%%%%
\begin{figure}[t]
    \mbox{\epsfig{file=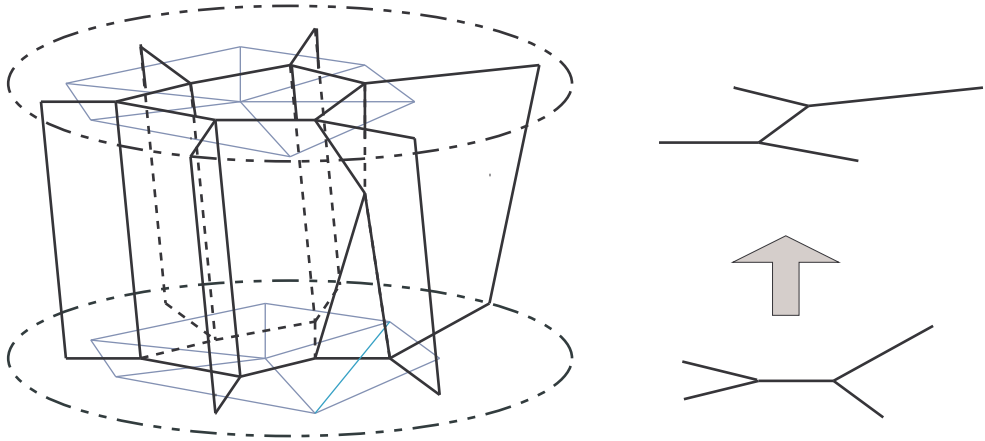}}
\caption{A potential first order contribution to the partition
function $Z_{RR}[s',s;\sigma;1]$ of eq.\ (3) and the corresponding
spin-network transition. A three dimensional situation is
reported. From the point of view of a 4 dimensional theory this
contribution correspond to a branched polyhedron dual to a singular
triangulation.}
\label{fig:spine}
\end{figure}
%%%%%%%%%%%%%%%%%%%%%%%%%%%%%%%%%%%%%%%%%%%%%%%%%%%%%%%%%%%%%%%%%%
In the previous equation, ${\cal H}$ is a diff-invariant 
definition of the Hamiltonian
constraints and $\sigma$ is a 2-dimensional branched surface embedded
over $\Sigma\times I[0,T]$. In figure \ref{fig:spine}
we reported an example of branched surface, and of the 
corresponding spin-network transition. The kind of surfaces that 
appear on the sum is explicitly determined by the 
kind of non null matrix elements of the super-Hamiltonian 
constraints. We want to emphasize, once more,
that this derivation is based on formal argument. The use of formal 
argument is of great help to understand the general picture
behind but one should stop here. The result of this formal 
derivation is that the functional integral associate
to quantum gravity is defined (or, better, should be) 
as a sum over branched surface interpolating between two spin 
networks. Moreover, it shows that one has to consider 
the sum over all the possible (not just one) such branched 
surfaces. This in a deep breakthrough toward the understanding
of loop quantum gravity and it shows a strict relationship
between loop quantum gravity and topological field theories of 
BF type opening the possibility of use technics developed for such theories.

The first occurrence, (to our knowledge) of the idea of formulating
the transition amplitude between two Hilbert space (spanned by
embedded colored graph) associated to two boundary component of a 
Manifolds, as a weight associated to the colored branched 
surface interpolending between the two graphs,
first appeared in Turaev and Viro \cite{Turaev:1992}. 
These authors used the terms: two dimensional polyhedral 
quantum field theory ({\it Spin foam} models are model 
of this kind). The next section will be
dedicated to the exposition of this idea in the case of 2+1
dimensional gravity or, more precisely, of the 
Ponzano-Regge-Turaev-Viro model.
  
%%%%%%%%%%%%%%%%%%%%%%%%%%%%%%%%%%%%%%%%%%%%%%%%%%%%%%%%%%%%%%%%%%
%%
%%           S E C T I O N
%%
%%%%%%%%%%%%%%%%%%%%%%%%%%%%%%%%%%%%%%%%%%%%%%%%%%%%%%%%%%%%%%%%%%

\section{The Ponzano-Regge-Turaev-Viro Model}

The Ponzano-Regge model and its quantum group analogous, the
Turaev-Viro partition function [it is known that its 
evaluation gives a quantum invariants of 3-manifolds],
are the better understood example of theory formulated 
on a triangulation $\Delta$ of a manifold $M$. 
{}From the physical point of view it is known (for details see
\cite{Ooguri:1991,Ooguri:1992a,Archer:1991}) that it correspond to the
quantization of Euclidean 2+1 dimensional gravity, i.e., to the
quantization of the $SU(2)$ BF theory in 3 dimension:
\begin{eqnarray}
Z_{BF}^{(3)}
  &=& \int [{\cal D}e(x)][{\cal D}\omega(x)]
             e^{{\rm i} \int {\rm Tr}[e\wedge R[\omega]]
                    +\Lambda~ {\rm Tr}[e\wedge e \wedge e] } 
\\
  &=& \int [{\cal D}A(x)][{\cal D}B(x)]
             e^{\frac{{\rm i}k}{4\pi} \int
          [{\rm Tr}[A\wedge F[A]]
          +\frac{2}{3} {\rm Tr}[A\wedge A\wedge A]] -
          [{\rm Tr}[B \wedge F[B]]
          +\frac{2}{3} {\rm Tr}[B\wedge B\wedge B]]  }
\nonumber \\
    &=& \left| \int [{\cal D}A(x)]
             e^{\frac{{\rm i}k}{4\pi} \int
           + {\rm Tr}[A\wedge F[A]]
          +\frac{2}{3} {\rm Tr}[A \wedge A \wedge A]}
     \right|^2
     = | Z_{CS} |^2
\nonumber
\end{eqnarray}
where
$$
 \omega  = \frac{1}{2}\left[A+B\right]    
~~~~~~~~
 e       = \frac{k}{8\pi}\left[A-B\right] 
~~~~~~~~
 \Lambda = \left(\frac{4\pi}{k}\right)^2
$$

The receipt for the construction 
of this model (to which we will refer to as the PRTV-model)
is the following. Color all the edge (1-dimensional
object of the triangulation) with irreducible representations of the
$SU(2)$ group and assign the following factor to the various 
object of the triangulation: 
1) to each edge the dimension of the representation
$\Delta_{c_e}$; 
2) to each triangle the inverse of the 3J symbol
made out by the three colors associated to each of its edges
$\theta(a_t,b_t,c_t)^{-1}$; 
3) to each tetrahedron the associated 6J symbol (a tetrahedron has 6 edges) 
${\rm Tet}[abed;cf]$. 

%%%%%%%%%%%%%%%%%%%%%%%%%%%%%%%%%%%%%%%%%%%%%%%%%%%%%%%%%%%%%%%%%%
\begin{figure}[t]
  \begin{center}
   \mbox{\epsfig{file=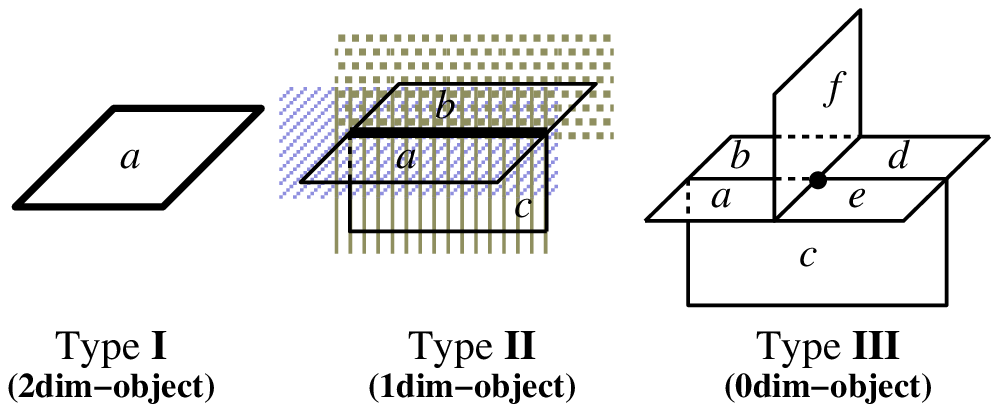,width=9cm}}
  ~\mbox{\epsfig{file=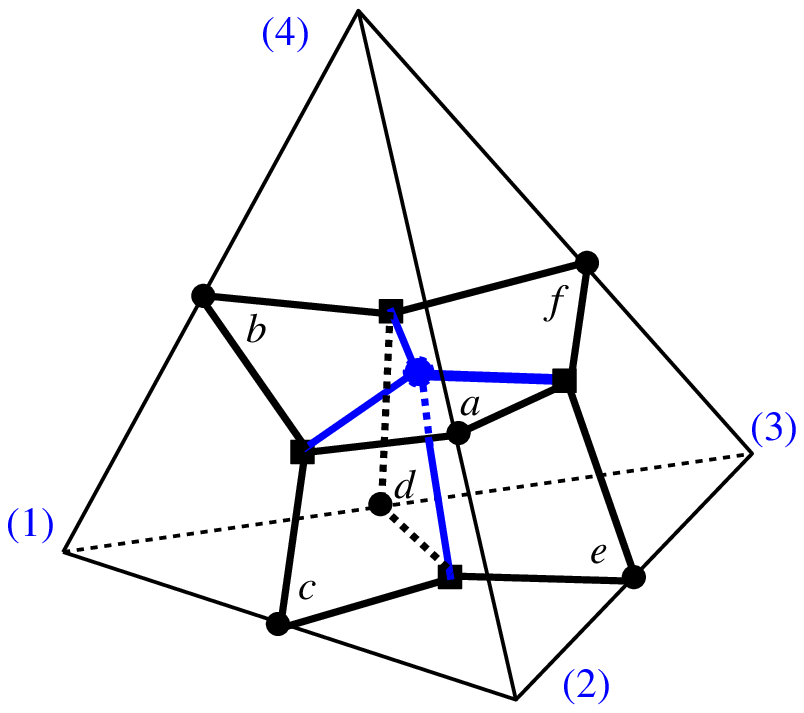,width=6cm}}
  \end{center}
\caption{The local description of the standard polyhedron dual to 
  a triangulation and a schematic representation of the duality.
  Note that in 3 dimension, a coloration of the edges of the triangulation 
  determine a unique coloration of the faces of its dual 2-skeleton,
  and {\it vice versa}. }
\label{duality}
\end{figure}
%%%%%%%%%%%%%%%%%%%%%%%%%%%%%%%%%%%%%%%%%%%%%%%%%%%%%%%%%%%%%%%%%%
In the case of the PRTV-model, it is useful to consider its dual
formulation, i.e. to take advantage of the duality  
between triangulation and standard 
polyhedron\footnote{A compact connected two dimensional polyhedron is
called quasi-standard if each point in it has a neighborhood
homeomorphic either to a plane (type I), to the union of three
half-planes with common boundary (type II), or the union of the four
half-plane dual to a tetrahedron in its barycenter (type III). We say
that it is standard if all its two dimensional component are disk.}
suggested by figure \ref{duality}.  
In the dual cellular complex $\Delta^*$ of the triangulation $\Delta$  we have
that to each $n$-dimensional cell of $\Delta$ is uniquely assigned a
3-$n$-dimensional cell of $\Delta^*$.  In three dimension, the objects
dual to an edge (1 dimensional object) is a surface (2=3-1 dimensional object).
This implies that to each coloration of the edges of the triangulation
$\Delta$ is associated a unique coloration of the face 
(two dimensional objects) of its dual standard polyhedron $\Delta^*$.
In turns, it is possible to consider the PRTV-model as a model
defined over a two dimensional branched surface. I.e., 
using this duality, we can interpret the PRTV model both as a
state sum model associated to  a triangulation but also has a state sum 
model associated to a branched surface:
\begin{equation}
TV_{G}[M^3,\Delta] = TV_G[{\cal S}_2^{\Delta^*}]
\label{ShadowSum}
\end{equation}
where by ${\cal S}_2^{\Delta^*}$ we denote the collection of
all the faces, edges and vertices (the 2-scheleton) of the cellular 
decomposition of $M$ dual to triangulation $\Delta$.
We will also use the notations ${\cal F}({\cal S}_2)$,
${\cal E}({\cal S}_2)$ and ${\cal V}({\cal S}_2)$
to denote the collection of its faces, edges and vertices, 
respectively.

This express the fact that the definition of the PRTV-model can be given
as a series of rules with respect to the objects of the
triangulation or as an equivalent set of rules with respect 
to the objects to its dual standard polyhedron:
\begin{center}
\begin{tabular}{|ll|l|cl|}
\hline
$\Delta$    &$\Delta^*$& & & 
Factor in $TV[M^3,\Delta]= TV_G[{\cal S}_2(\Delta^*)]$ \\
\hline
            &          &               &  & \\
vertex       & 3-cell   & Type 0        &  
            & $\Lambda^{-1} = 
              \left[\sum_a \Delta_a  \right]^{(-1)}$ \\[3mm] 
edge        & face     & Type {\bf I}  &~~
            & $\Delta_a = dim(a) = \DELTA{a}$ \\[3mm]
face        & edge     & Type {\bf II} &
            & $\theta(a,b,c)^{-1} = 
              \left[{\THETA{a}{b}{c}} \right]^{(-1)}$ \\[3mm]
tetrahedron & vertex   & Type {\bf III} &
            & ${\rm Tet}[{abed};{cf}]=\TET{a}{b}{f}{e}{d}{c}$ \\
\hline
\end{tabular}
\end{center}
One of the nice characteristic of the dual formulation
is that it involves only the coloring of a 2-dimensional
standard polyhedron embedded into the Manifold. Completing
the transition to the dual world it is possible to
to forget the original triangulation and to directly associate a 
state-sum to any standard polyhedron of the ($M^3$) manifold.
\begin{equation}
TV_{G}[{\cal S}_2]
    =  \Lambda^{\scriptscriptstyle -\#(3-cells)} \!\!\!\!\!
\sum_{\vec{c}\in D({\cal S}_2,G)}
    \left[\!\prod_{f\in {\cal F}({\cal S}_2)}\! 
    \left(\DELTA{c_f}\right)^{\chi(f)}  
    \right]
    \left[\!\prod_{e\in {\cal E}({\cal S}_2)}\!
    \left({\THETA{a_e}{b_e}{c_e}}\right)^{-\chi(e)}
    \right]
    \left[\!\prod_{v\in {\cal V}({\cal S}_2)}\!
    \TET{a_v}{b_v}{f_v}{d_v}{e_v}{c_v} 
    \right]
\label{def:TV1}
\end{equation}
where $\chi(f)$ and $\chi(e)$ are the Euler characteristics of the
face $f$ and of the edge $e$, respectively. For the purpose of 
the present discussion the introduction of the Euler characteristics
is completely irrelevant. All the edges and faces of a standard
2-polyhedron dual to a triangulation of $M^3$ have
Euler-characteristic equal to 1! We explicitly wrote  
the Euler characteristic because one of the interesting characteristic
of the PRTV partition function is that is is defined for any 
stratification of the manifolds $M$ whose two skeleton is a standard
polyhedron. That means that it is defined on the larger class of
singular triangulations \cite{Turaev:1992}. For the rest of the 
discussion we will always deal with regular polyhedron dual to 
triangulation and we will always replace the Euler characteristic 
with the appropriate value for polyhedrons belonging to the dual 
of a triangulation: 1. 

This model manifest its topological nature from the fact that it is 
possible to prove that $TV_{G}[{\cal S}_2]$ it is independ
by the chosen standard polyhedron dual to a triangulation 
(pseudo-triangulation) of $M$. Infact, it is possible to prove 
that, if ${\cal S}_2$ and $\tilde{\cal S}_2$ are two different standard
polyhedron dual to triangulations of the same manifold $M$, then:
\begin{equation}  
TV_{G}[M]=TV_{G}[{\cal S}_2]=TV_{G}[\tilde{\cal S}_2]
\label{Tnature}
\end{equation}  

\subsection{Decorating a 2 dimensional standard polyhedron
(spin foam)}

In the previous section we used the notation $D({\cal S}_2,G)$ to
denote all the possible decorations (coloring) of the 2-surfaces of a
standard polyhedron with inequivalent irreducible representations 
of a group $G$.
The admissibility condition amount to the existence of an 
intertwining matrix (in the case of the $SU(2)$ group, if it exists, 
it is unique up to normalization) between the three representation 
associated to the three faces that meet at an edge (a type II object
of figure \ref{duality}). It is natural to extend this concept
of decoration to the more general situations where there are 
n-faces that meet at a given edge. In this case, we have to assign
to each edge an invariant contractor on the tensor product
of the spaces of all the representations associated to the faces 
that meet at the edge.
In this way, we have a decoration of the two- and one-imensional objects
of the 2-polyhedron completly analogous to the  decoration 
(coloring) of the edges and vertices of a graph $\gamma$ that we 
used to label the spin networks  basis of equation
(\ref{eq:defSPINnet})
that, using the notation $D({\cal S}_2,G)$, can be rewrite
as:
\begin{equation}
f_\gamma(g_{e_1},\ldots,g_{e_n}) = 
\!\sum_{\vec{\pi},\vec{\iota} \in D(\gamma,SU(2))} \!
   f(\gamma,\vec{\pi},\vec{\iota}) 
  ~~{\cal T}_{\gamma,\vec{\pi},\vec{\iota}~}[A]
.
\label{eq:decSPINnet} \end{equation}

This completes the definition of the concept of a group decorated 2
dimensional oriented polyhedron. This extension to non simple
polyhedra of this concept (implicetly contained in
\cite{Reisenberger:1994}) of group decoratation of 2-polyhedron with 
boundary is due to Baez. He used, the now
generaly accepted, term  {\it spin foam} to denote 
a single group decorated 2-polyhedron and the
term {\it spin foam model} to denote a theory based on the sum on all
the possible decoration (coloration). We refer the reader to the 
Baez work \cite{Baez:1998} for more details and the discussion
of the additional care needed on the case of a general group.

%%%%%%%%%%%%%%%%%%%%%%%%%%%%%%%%%%%%%%%%%%%%%%%%%%%%%
%%
%%
%%
%%%%%%%%%%%%%%%%%%%%%%%%%%%%%%%%%%%%%%%%%%%%%%%%%%%%

\subsection{The Hilbert space on the Boundary
and the functorial nature of the invariant}

Following Turaev and Viro \cite{Turaev:1992}, we note that it is
possibile to generalize the PRTV-model to include the case 
of a manifold $M$ with boundary with a decorated (colored)
graph sitting on its boundary $\partial M$. The PRTV
partition function (\ref{def:TV1}) in the case of  a decorated
polyhedron ${\cal S}_2$ with a graph $\gamma$ 
on the boundary (i.e., $\partial{\cal S}_2={\cal S}_2
\cap M^3 = \gamma$) is defined as 
\begin{eqnarray}
TV_G[{\cal S}_2;(\gamma,\vec{c})] \!\!
    &=& \frac{1}{N(\gamma,\vec{c})} 
        \Lambda^{\frac{\#(2-cells~of~\gamma)}{2}}
        \Lambda^{\scriptscriptstyle -\#(3-cells)}
\times \label{def:TV2}\\
 & & ~~\times \sum_{\vec{c}\in D_{in}{({\cal S}_2,G)}}
    \bigg[\prod_{f\in {\cal F}_{in}({\cal S}_2)} \!\!\!
          \left(\DELTA{c_f}\right)   \bigg]
    \bigg[\prod_{e\in {\cal E}_{in}({\cal S}_2)} \!\!\!
    \left({\THETA{a_e}{b_e}{c_e}}\right)^{-1}
    \bigg]
    \bigg[\prod_{v\in {\cal V}({\cal S}_2)} \!\!\!
          \TET{a_v}{b_v}{f_v}{d_v}{e_v}{c_v} \bigg]
\nonumber \\
\frac{1}{N(\gamma,\vec{c})} 
&=& \sqrt{
       \prod_{e\in {\cal E}(\gamma)}\DELTA{c_e}
       \prod_{v\in {\cal V}(\gamma)}
          \left( \THETA{a_v}{b_v}{c_v}\right)^{-1}
    }
\label{SNnorm}
\end{eqnarray}
where with ${\cal F}_{in}({\cal S}_2)$ and 
${\cal E}_{in}({\cal S}_2)$ we have denoted the
faces and edges of ${\cal S}_2$ that do not end
or belong to the boundary of ${\cal S}_2$. Moreover
$D_{in}{({\cal S}_2,G)}$ denotes a decoration of the
internal faces ${\cal F}_{in}({\cal S}_2)$ compatible
with the decoration of the graph $\gamma$ on the 
boundary\footnote{A similar costruction was used by  
G. Carbone, M. Carfora and  A. Marzuoli to construct a true 
invariant for manifold with non empty 
boundary \cite{Carfora:1998}}
($c_f=c_e$ if $f\in{\cal F}({\cal S}_2)$, 
$e\in{\cal E}(\gamma)$ and $e\subset \partial f$).
.

%%%%%%%%%%%%%%%%%%%%%%%%%%%%%%%%%%%%%%%%%%%%%%%%%%%%%%%%%%%%%%%%%%
\begin{figure}[t]
  \begin{center}
     \mbox{\epsfig{file=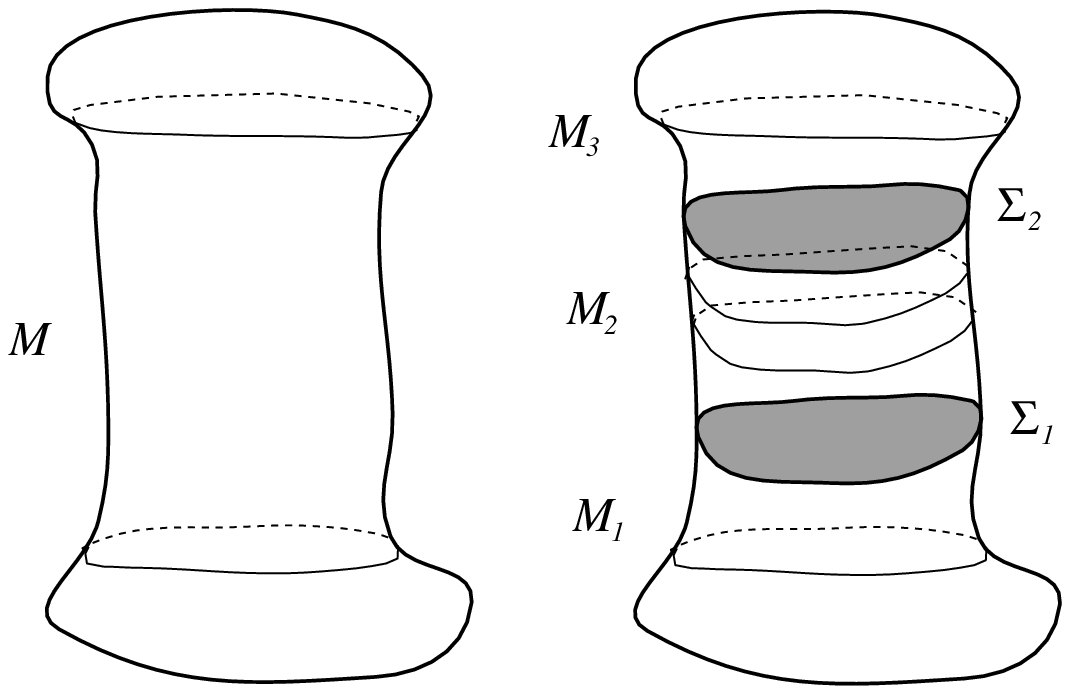,width=8cm}}
 ~~~~\mbox{\epsfig{file=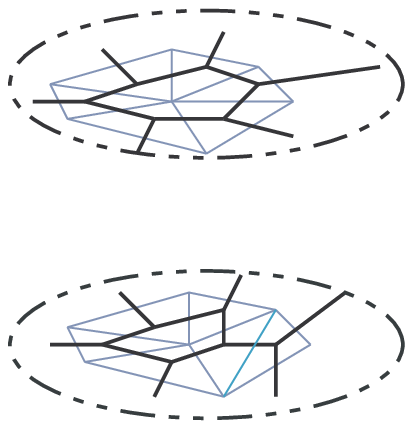}}
  \end{center}
\caption{A possible decomposition of an orientable 
Manifold in three components.  This drawing has only a schematic 
purpose since we suppose that $\partial M_3 = - \Sigma_2$, 
$\partial M_1 = \Sigma_1$ and $\partial M_2 = \Sigma_2 \cup - \Sigma_1$.}
\label{3pieces}
\end{figure}
%%%%%%%%%%%%%%%%%%%%%%%%%%%%%%%%%%%%%%%%%%%%%%%%%%%%%%%%%%%%%%%%%%

Using the previous generalization of the PRTV-model we can 
reinterpred the state sum model of equation (\ref{ShadowSum})
in the following way. As suggest by figure \ref{3pieces} we 
can imagine to split the manifold
$M$ in three disjoint components glued together along the
common boundary and rewrite the partition function associate
to any given triangulation as:
\begin{eqnarray}
TV_G[M^3,\Delta]
   &=& \sum_{\vec{c}'\in D(\gamma,G)} ~ \sum_{\vec{c}\in D(\gamma,G)}
        TV_G[{\cal S}_2^{\Delta_1^*};(\gamma',c')]
        \frac{1}{N(\gamma',c')^2}
\nonumber \\ & & \qquad
        TV_G[{\cal S}_2^{\Delta_2^*};(\gamma'\cup\gamma, c+c' )]
        \frac{1}{N(\gamma,c)^2}
        TV_G[{\cal S}_2^{\Delta_3^*};(\gamma,c)]
\label{split} 
\end{eqnarray}
where $\gamma\in\Sigma_1$ and $\gamma'\in\Sigma_2$ are the graphs
determined by the intersection of the 2 bone of the dual cellular
decomposition $\Delta^*$ with $\Sigma_1$ and $\Sigma_2$, respectively.
Now, getting the intersection of $\Delta^*$ with a 2 surface
$\Sigma$ we will have that the 3 cells are mapped into 2 cells,
the 2 cells are mapped into 1-cells (the edge of $\gamma$) and
the 1 cells into vertices of $\gamma$. In this operation
we are indeed inducing a coloring of $\gamma$ from the coloring
of the branched polyhedron. Moreover, if $\Delta$ is a triangulation
of $M$ then the resulting graph is trivalent.
Equation (\ref{split}) has a nice interpretation in terms of an
operator equation. Consider the Hilbert spaces ${\cal H}_{\gamma}$ 
of section 2.1 associated to graphs embedded on the two Cauchy surfaces 
$\Sigma_1$ (${\cal H}_{\gamma_1}$) and $\Sigma_2$ (${\cal H}_{\gamma_2}$).
In the normalization of \cite{DePietri:1996} and
\cite{DePietri:1997} the completeness of  ${\cal H}_{\gamma}$
is given by  
\begin{equation}
  \hat{1}_{{\gamma}}
  = \sum_{\vec{c}\in D(\gamma,G)} \frac{1}{N(\gamma,\vec{c})^2} ~
       |\Sigma;\gamma,\vec{c}\rangle ~\langle \Sigma;\gamma,{\vec{c}}|
 ~~~,
\end{equation}
Moreover, we can define an operator $\hat{P}_{\gamma_2,\gamma_1}^{(M_2)}$
\begin{eqnarray}
\hat{P}_{\gamma_2,\gamma_1}^{M_2}
   &:& {\cal H}_{\gamma_1} \Longrightarrow {\cal H}_{\gamma_2}
\\[1mm]
\langle \gamma',c' |\hat{P}_{\gamma',\gamma}^{M_2} |\gamma,c\rangle 
&=&  TV_G[{\cal S}_2^{\Delta_2^*};(\gamma'\cup\gamma, c+c' )].
\label{DEF1}
\end{eqnarray}
This operator $\hat{P}_{\gamma',\gamma}^{M_2}$ does not
depends on the interior of the simple polyhedron 
${\cal S}_2^{\Delta_2^*}$ used to compute the transition amplitude
\cite{Turaev:1992}. Using this notation we can rewrite 
write equation (\ref{split}) as
\begin{equation}
  TV_G[M^3,\Delta] = \langle \emptyset |
          \hat{P}_{\emptyset,\gamma_2 }^{M_3}
          \hat{P}_{\gamma_2 ,\gamma_1 }^{M_2}
          \hat{P}_{\gamma_1 ,\emptyset}^{M_1}
                  |\emptyset\rangle ~~.
\end{equation}
In the previous notation we have implicitly assumed that the boundary
of each of the manifold $M_i$ can be naturally seen as made of two
disjoint components ($\partial M = (\partial M)^{+} \cap (\partial
M)^{-}$) one associated to the {\it future} $(\partial M)^{+}$ and the
other to the {\it past} $(\partial M)^{-}$.  
$\hat{P}_{\gamma',\gamma}^{M}$ is the operator of equation 
(\ref{DEF1}) between the Hilbert space associate to the graph 
$\gamma\subset -(\partial M)^{-}$ and the Hilbert space 
associated to the graph $\gamma'\subset (\partial M)^{+}$.  With
this convention it is natural to define the composition of two 
manifolds with boundary $M_1$ and $M_2$ when $(\partial M)_2^{+}=-(\partial
M)_1^{-}$ as the manifold $M=M_2\circ M1$ that one obtains gluing
$M_1$ and $M_2$ along the isomorphic boundary $(\partial
M)_2^{+}=-(\partial M)_1^{-}$.  At this point we are ready to discus
the functorial nature of the invariant. From the independence 
of the PRTV state sum from the chosen polyhedron
we have that the $\hat{P}$ define a semifunctor
on the space of manifolds with a fixed graph on their boundary:
\begin{equation}
 \hat{P}_{\gamma' ,\gamma}^{M_2\circ M_1} = \hat{P}_{\gamma'
 ,\gamma''}^{M_2}\hat{P}_{\gamma'',\gamma}^{M_1}
\end{equation} 
This $\hat{P}$ does not properly define a topological field theory
since it is not a functor 
($\hat{P}_{\gamma,\gamma}^{\Sigma\times I} \neq 1_{\gamma}$).  
There is
however a standard procedure that allows to define a functor from a
semifunctor. In fact, if we consider the Hilbert space
$\overline{{\cal H}_\gamma} ={\cal
H}_\gamma/Ker(\hat{P}_{\gamma,\gamma}^{\Sigma\times I})$ from
$\hat{P}_{\gamma',\gamma}^{M}=\hat{P}_{\gamma',\gamma}^{M}\circ
\hat{P}_{\gamma,\gamma}^{\Sigma\times I}$ follows that its restriction
to the family of Hilbert spaces $\overline{{\cal H}_\gamma}$ become a
functor. I.e, we used the operator 
$\hat{P}_{\gamma,\gamma}^{\Sigma\times I}$
as the projector on the physical Hilbert space of the
theory. In this way the PRTV-model defines a 2+1 dimensional 
quantum field theory \cite{Turaev:1992} in the Atiyah terminology
\cite{Atiyah:1989}.   

{}From the physical point of view this discussion can be
seen as a computation of Hartle-Hawking wave function and of the
topology change amplitude of the 3-dimensional lattice gravity
of Ponzano and Regge \cite{Ooguri:1992a}.
In consideration of a Hamiltonian (initial values)
formulation in terms of loop-variable this is particularly nice.
In fact, in dimension $n$ the Hamiltonian formulations is defined on an
$(n-1)$ dimensional Cauchy surface and in any dimension the intersection
of an $(n-1)$ surface and any 2-dimensional shadow is an 1-dimensional
graph embedded on the Cauchy surface. Even more interesting is the 
fact the Hilbert space structure associated to the BF theory is 
very similar to the one of loop quantum gravity.

%%%%%%%%%%%%%%%%%%%%%%%%%%%%%%%%%%%%%%%%%%%%%%%%%%%%%%%%%%%%%%%%%%
%%
%%           S E C T I O N
%%
%%%%%%%%%%%%%%%%%%%%%%%%%%%%%%%%%%%%%%%%%%%%%%%%%%%%%%%%%%%%%%%%%%

\section{Four dimensional models}

In previous section we showed how the Ponzano-Regge model
correspond to the BF theory in 3 dimension. In this
section we discuss an example of 4 dimensional theory that 
can be expressed within the spin foam formalism: the 
Crane-Yetter-Ooguri model of BF theory and the spin foam 
model of Barret and Crane. We do not discus here the 
model of Reisenberger \cite{Reisenberger:1997a,Reisenberger:1997b}. 
All this models share the characteristic
to be based on a theory of the BF type.  This lead Freidel and 
Krasnov \cite{Freidel:1998} to propose a way to systematicaly derive
the {\it spin foam model} associated to any theory that admits a formulation 
in terms of  an action of BF type plus Lagrange's multipliers imposed 
constraints. 

%%%%%%%%%%%%%%%%%%%%%%%%%%%%%%%%%%%%%%%%%%%%%%%%%%%%%%%%%%%%%%%
\begin{figure}[t]
\begin{center}
\mbox{\epsfig{file=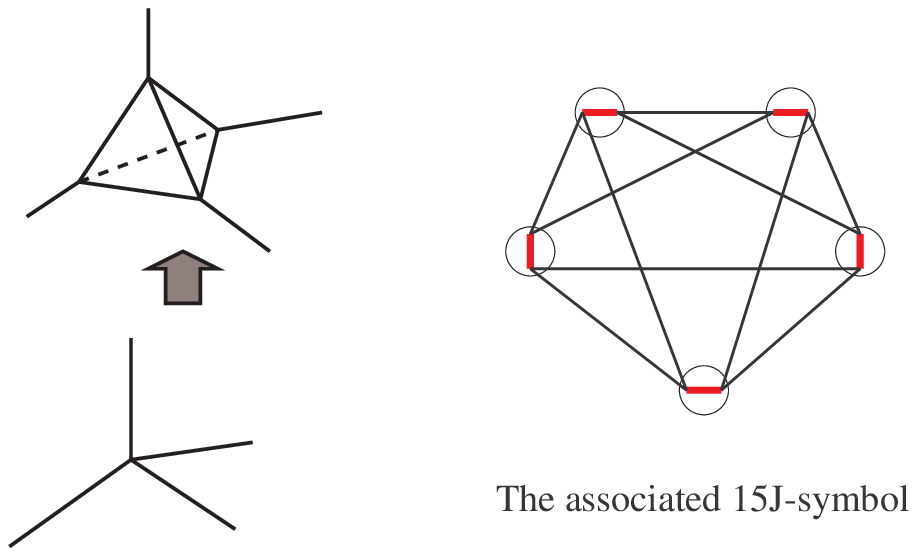}} 
\end{center}
\caption{The elementary vertex of the BF theory. There are 5 nodes
(with the associated intertwiner) and 10 links (with the associated
representation). To the elementary vertex are indeed associated 
15 colors.}
\label{fig:vertexBF}
\end{figure}
%%%%%%%%%%%%%%%%%%%%%%%%%%%%%%%%%%%%%%%%%%%%%%%%%%%%%%%%%%%%%%%%%%

\subsection{The Crane-Yetter-Ooguri model}

The state sum formulation of the topological $BF$ theory in
4-dimensions 
\begin{eqnarray}
Z_{BF}^{(4)} 
   &=& \int [{\cal D}A(x)][{\cal D}B(x)]
           e^{{\rm i}\int {\rm Tr}[B \wedge F[A]] 
               + \Lambda~ {\rm Tr}[B\wedge B] }
\label{eq:ActionBF}\\
   &=& \int [{\cal D}A(x)]
           e^{\frac{{\rm i}}{2\Lambda} \int{\rm Tr}[F[A] \wedge F[A]]}
%%%\nonumber\\ &=& 
   = \int [{\cal D}A(x)]
        e^{\frac{{\rm i}}{2\Lambda} \int d\big(
        {\rm Tr}[A \wedge F[A]]+\frac{2}{3}{\rm Tr}[A\wedge A\wedge A]
        \big)}
\nonumber
\end{eqnarray}
with cosmological constant is given by the Crane-Yetter
\cite{Crane:1997} or by the Ooguri (without cosmological constant)
\cite{Ooguri:1992b} models.  The partition function of the model
associated to a triangulation $\Delta$ of a manifold $M$ can be
written as~:
\begin{equation}
Z_{BF}(\Delta, \Lambda) = \sum_{j_f,i_t} \prod_{f} dim_q(j_f)
\prod_{v} \phi_{q,v}(\vec \jmath, \vec \imath),
\end{equation}
where $q=\exp ({\rm i}\Lambda)$, $j_f$ denotes a coloring of the faces
of $\Delta$ by irreducible representation of $U_q(su(2))$, $i_t$
denotes a coloring of the tetrahedra of $\Delta$ by intertwiners and
the sum is over all such colorings with $j<{ \pi \over\Lambda}$.
Moreover $dim_q(j)$ denotes the quantum dimension of the
representation of spin $j$ and $\phi_v(\vec \jmath, \vec \imath)$
denote the quantum $15$-j symbol associated with the 4-simplex $v$.
More precisely, associated to a 4-simplex $v$ we consider a graph
$\Gamma_v$ given by the intersection between the 2-skeleton of the
complex which is dual to the triangulation and the boundary of the
4-simplex $v$ (see figure \ref{fig:vertexBF}).  $\Gamma_v$ corresponds
to the pentagram graph and we color its 10 edges by $\vec j$ and its 5
vertices by $\vec i$.  $\phi_{q,v}(\vec \jmath, \vec \imath)$
corresponds to the Reshetikhin-Turaev evaluation of the colored graph
$\Gamma_v$.  This state sum does not depend on the triangulation when
correctly normalized by a factor $N^{n_0-n_1}$, where $n_0,n_1$ is the
number of $0,1 $ simplices, and it corresponds to the evaluation of
the BF partition function of equation (\ref{eq:ActionBF}).
 
\subsection{The Barret-Crane model}

The Relativistic-Spin-Foam model is a modification of the state sum
(\ref{eq:ActionBF}) in the case of the $so(4)$ gauge group in which the
conditions corresponding to the quantum 4-simplex conditions (when
$\Lambda=0$) are imposed by hand. Yetter generalized this construction
to the case of non vanishing $\Lambda$ \cite{Yetter:1998}.  
Yetter emphasized that the quantum group that should be considered 
is $U_q(su(2)) \otimes U_{q^{-1}}(su(2))$.  This state sum model can 
be written~:
\begin{eqnarray}\label{BC}
Z_{BC}(\Delta, \Lambda)
  &=& \sum_{j_f,\jmath^\prime_f,i_t, \imath^\prime_t } \prod_{f}
      dim_q(j_f) dim_{q^{-1}}(\jmath^\prime_f)
%%%##%%% \\ &&~~~~~
\prod_{v}  \phi_{q,v}(\vec \jmath, \vec \imath)
      \phi_{q^{-1},v}(\vec{\jmath^{\prime}}, {\vec {\imath^\prime}})
      \delta_{\vec \jmath, \vec{\jmath^{\prime}}}\delta_{\vec \imath,
      \vec{\imath^\prime }}.
%%%##%%%\nonumber
\end{eqnarray}
So this model corresponds to two copies of $su(2)$ $BF$ theories (or
an $so(4)$ $BF$ theory) together with $15$ constraints imposed for
each 4-simplex.  The presence of $\delta_{\vec j, \vec {\jmath^\prime
}}$ and $\delta_{\vec i,\vec{\imath^\prime }}$ corresponds to 
the geometricity constraint. In \cite{DePietri:1998} it is 
shown that the above spin foam model can be associated to a well
defined field action: the $so(4)$ Plebanski action \cite{Plebanski:1977}. 
This is a deformed BF theory depending on an $so(4)$ connection $\omega=
\omega_\mu^{IJ}X_{IJ} dx^\mu$, a two form valued into $so(4)$ $B=
B^{IJ}_{\mu\nu} X_{IJ} dx^\mu\wedge dx^\nu$, and a scalar symmetric
traceless matrix $\phi_{[IJ][KL]}$ ($\epsilon^{IJKL} \phi_{IJKL}=0$)
that acts as Lagrange multiplier fields. This action reads:
\begin{eqnarray}
  {\cal S}[\omega;B;\phi] &=& \int_M\!\!\!
  \bigg[ B^{IJ}\wedge F_{IJ}(\omega)
- \frac{\Lambda}{4}\epsilon_{IJKL} B^{IJ} \wedge B^{KL}
%%%##%%% \nonumber \\  & &~~
- \frac{1}{2} \phi(B)_{IJ}\wedge B^{IJ}\bigg] ~,
\label{AP}
\end{eqnarray}
where $\phi(B)_{IJ}= \phi_{IJKL} B^{KL}$. In this context, the 
geometry constraints are imposed, at the quantum level, by the 
integration over Lagrange multiplier fields $\phi_{[IJ][KL]}$.
This action has two nice characteristics: (1) one of its sectors of
solution is closely connected to gravity; (2) it is a deformation
of a quantizable topological theory (The BF theory) that can
play a role analogous to the one of free theory in standard
quantum field theory. Moreover, it is intriguing that this 
description of gravity as a constraint BF theory can be 
generalized to higher dimension \cite{Freidel:1999b}. 

%%%%%%%%%%%%%%%%%%%%%%%%%%%%%%%%%%%%%%%%%%%%%%%%%%%%%%%%%%%%%%%%%%
%%
%%           S E C T I O N
%%
%%%%%%%%%%%%%%%%%%%%%%%%%%%%%%%%%%%%%%%%%%%%%%%%%%%%%%%%%%%%%%%%%%

\section{Conclusion}

In this work we showed that the spin foam formalism gives
an unified framework for the discussion of an important
class of diff-invariant quantum theories. Between them: topological
BF theory in 3 and 4 dimension, the four dimensional
formulation of loop quantization of gravity,
the Barret-Crane model. The main unsolved problems of the 
general picture we described here are:
\begin{enumerate}
\item The state sum model we described are properly defined only on 
      a single given triangulation (or its dual spin-foam). If this 
      is fine in the case of a topological field theory without local
      degree of freedom this is questionable in the case of gravity.
      Work is now in progress \cite{DePietri:1999} on the direction
      of dealing with all the possible triangulation.
      This is somehow analogue to consider generalization (where
      the triangulation is not the only dynamical variable) of the
      model of dynamical triangulations \cite{Ambjorn:1997}.
\item The connection between the spin foam models and the exponential
      of the constraints need to be improved with a rigorous 
      mathematical analysis of the derivation. 
\item A systematical study of the physical implication of taking different
      weight factor is missing.   
\end{enumerate}

%%%%%%%%%%%%%%%%%%%%%%%%%%%%%%%%%%%%%%%%%%%%%%%%%%%%%
%%
%%
%%
%%%%%%%%%%%%%%%%%%%%%%%%%%%%%%%%%%%%%%%%%%%%%%%%%%%%
\vspace{.5cm}
\noindent {\sc Acknowledgment:}\\
This work was greatly influenced by numerous discussion with
M. Carfora, L. Freidel, J. Lewandowski, L. Lusanna, A. Marzuoli,
M. Pauri, C. Rovelli and J. Zapata, to whom all I'm extremely grateful.
The work of R.D.P. at CPT Marseille is supported by a Dalla Riccia
Fellowship.

\vspace{1.0cm}

%% =============================================================
%%   Bibliography:
%% =============================================================

%% \bibliographystyle{amsalpha}
%% \bibliography{loop}
%% \end{document}

%% =============================================================
%%   Bibliography: II
%% =============================================================
\newcommand{\etalchar}[1]{$^{#1}$}
\providecommand{\bysame}{\leavevmode\hbox to3em{\hrulefill}\thinspace}

\end{document}